\documentclass[aps,prb,citeautoscript,twocolumn,showpacs,superscriptaddress]{revtex4}  
\usepackage[pdftex]{graphicx}  
\usepackage{graphicx}  
\usepackage{epsfig}
\usepackage{dcolumn}   
\usepackage{bm}        
\usepackage{amssymb}   
\usepackage{amsmath}
\usepackage[latin1]{inputenc}
\usepackage[usenames,dvipsnames]{color}
\usepackage{color}      
\usepackage{setspace}
\usepackage{subfigure}  

\usepackage{ulem}  

\begin{document}
\title{Bichromatic dressing of a quantum dot detected by a remote second quantum dot}
\author{M. Maragkou}
\affiliation{Departamento de F{\'i}sica de Materiales, Universidad Aut{\'o}noma de Madrid, 28049, Spain.}
\author{C. S{\'a}nchez-Mu{\~n}oz}
\affiliation{Departamento de F{\'i}sica Te{\'o}rica de la Materia Condensada, Universidad Aut{\'o}noma de Madrid, 28049, Spain.}
\author{S. Lazi{\'c}}
\author{E. Chernysheva}
\affiliation{Departamento de F{\'i}sica de Materiales, Universidad Aut{\'o}noma de Madrid, 28049, Spain.}
\author{H.~P. van der Meulen}
\affiliation{Departamento de F{\'i}sica de Materiales, Universidad Aut{\'o}noma de Madrid, 28049, Spain.}
\affiliation{Unidad Asociada CSIC-SEMICUAM.}
\affiliation{Instituto de Ciencia de Materiales 'Nicol{\'a}s Cabrera', Universidad Aut{\'o}noma de Madrid, 28049, Spain.}
\author{A. Gonz{\'a}lez-Tudela}
\affiliation{Departamento de F{\'i}sica Te{\'o}rica de la Materia Condensada, Universidad Aut{\'o}noma de Madrid, 28049, Spain.}
\author{C. Tejedor}
\affiliation{Departamento de F{\'i}sica Te{\'o}rica de la Materia Condensada, Universidad Aut{\'o}noma de Madrid, 28049, Spain.}
\affiliation{Unidad Asociada CSIC-SEMICUAM.}
\affiliation{Instituto de Ciencia de Materiales 'Nicol{\'a}s Cabrera', Universidad Aut{\'o}noma de Madrid, 28049, Spain.}
\affiliation{Condensed Matter Physics Center (IFIMAC), Universidad Aut{\'o}noma de Madrid, 28049, Spain.}
\author{L.~J. Mart{\'i}nez}
\author{I. Prieto}
\author{P.~A. Postigo}
\affiliation{Instituto de Microelectr{\'o}nica de Madrid, Centro Nacional de Microelectr{\'o}nica, Consejo Superior de Investigaciones
Cient\'ificas, Isaac Newton 8, PTM Tres Cantos, E-28760 Madrid, Spain.}
\author{J.~M. Calleja}\email{jose.calleja@uam.es} 
\affiliation{Departamento de F{\'i}sica de Materiales, Universidad Aut{\'o}noma de Madrid, 28049, Spain.}
\affiliation{Unidad Asociada CSIC-SEMICUAM.}
\affiliation{Instituto de Ciencia de Materiales 'Nicol{\'a}s Cabrera', Universidad Aut{\'o}noma de Madrid, 28049, Spain.}
\affiliation{Condensed Matter Physics Center (IFIMAC), Universidad Aut{\'o}noma de Madrid, 28049, Spain.}

\date{\today}

\begin{abstract}

We demonstrate an information transfer mechanism between two dissimilar remote InAs/GaAs quantum dots weakly coupled to a common photonic crystal microcavity. Bichromatic excitation in the s-state of one of the dots leads to the formation of dressed states due to the coherent coupling to the laser field, in resonance with the quantum dot. Information on the resulting dressed structure is read out through the photo-luminescence spectrum of the other quantum dot, as well as the cavity mode. The effect is also observed upon exchange of the excitation and detection quantum dots. This quantum dot inter-talk is interpreted in terms of a cavity-mediated coupling involving acoustic phonons. A master equation for a three level system coherently pumped by the two lasers quantitatively describes the behavior of our system.
Our result presents an important step towards scalable solid-state quantum networking based on coupled multi-quantum-dot-cavity systems, without the need of using identical quantum emitters.

\end{abstract}


\maketitle
\section{Introduction}

The control of the light-matter interaction at the nanometer scale and its use for the development of novel schemes
for processing of quantum information have been among the most active research areas in the last years. Strong light-matter interaction is often achieved in two-level systems in the realm of cavity quantum electrodynamics (QED) \cite{QED1}. Cavity-QED experiments in semiconductor quantum dots (QDs) \cite{QEDdots2}  gave rise to a number of applications, particularly in quantum information processing and quantum networking \cite{QIP3, QNetw3}, as well as for single photon emitters \cite{SPS1,SPS2,SPS3}. This is partly due to the simultaneous confinement of excitons and photons \cite{QEDdots2,2,3,4,5} in these systems. If the two-level system is coherently driven by strong resonant excitation, mixed exciton-photon states (dressed states) are formed, which are at the origin of the well-known Mollow triplet \cite{6} observed in the optical emission of atoms \cite{10} and QDs under resonant excitation \cite{7,8,9}. Strong cavity emission also occurs even when the QD emitter is not in resonance with the cavity \cite{3}, and several cavity-feeding mechanisms have been proposed \cite{4,5} for this phenomenon, including the intermediation by acoustic phonons \cite{11}. Simultaneous coupling of more than one QD to the same cavity mode (CM) is also possible when there is sufficient spectral and spatial overlap \cite{13,14,15,16}. In principle, even under non resonant conditions, the coupling between multiple QDs and the cavity could allow to store and retrieve information on the coupled system through different spectral channels, opening the way to transferring quantum information via photons between remote nodes of a solid-state-based network. Preliminar steps in this direction, as cavity-mediated QD coupling between two QDs coupled to the same cavity, have been reported for p-state excitation by E.Gallardo et al \cite{15}. and for s-state excitation by A.Majumdar et al \cite{16}. Also dressing of a QD state by a laser field and readout of the dressed spectral distribution by the cavity emission has been reported by A.Majumdar et al.   \cite{12}

In this work we study a system of two distant InAs/GaAs QDs weakly coupled to a common CM under coherent bichromatic excitation. We show that dressing of one of the QDs by the laser field can be effectively readout by the optical emission of a second QD, in addition to the cavity emission. This result is a significant advance over previous work \cite{12} in the use of QDs for solid state quantum networks, as it demonstrates the feasibility of information transfer between distant QDs coupled to a common cavity, bringing closer the use of QD/cavity pairs as nodes of a network for quantum information processing. Upon simultaneous pumping of the system by two continuous wave (CW) lasers, dressed states are created. One (fixed laser) is in resonance with the s-state of one of the QDs  while the other laser (variable laser) continuously scans across a small energy range around the frequency of the fixed laser. Because the intensities of both (fixed and variable) lasers are comparable, the variable laser cannot be treated as a linear perturbation of the fixed one, i.e. our measurement is not a typical pump-probe experiment. The spectral distribution of the dressed states in the QD excited by the laser field determines the population of the excited state of the other QD, which is measured by the intensity variation of its optical emission as well as of the cavity emission as a function of the detuning of the variable laser. Information on the dressed states of one QD is, therefore, obtained from the emission intensity of the other QD, also weakly coupled to the same CM. The excitation and detection ports are interchangeable. Indeed we show that the cavity mediated inter-talk between QDs operates in both directions: up-conversion (UC) for excitation at lower energy than that of detection and down-conversion (DC) for the opposite way. 
The coupling requires the exchange (emission for DC or absorption for UC) of acoustic phonons. The dependence of measurable properties on both fixed and variable laser frequencies and intensities are described by a master equation for the dynamics of the system density matrix $\rho$. The experimental results can be successfully fitted by the calculated intensity distributions, thus giving a good understanding of the physics behind the experiments.

\section{Theory}

\begin{figure}
\includegraphics[width=0.4\textwidth]{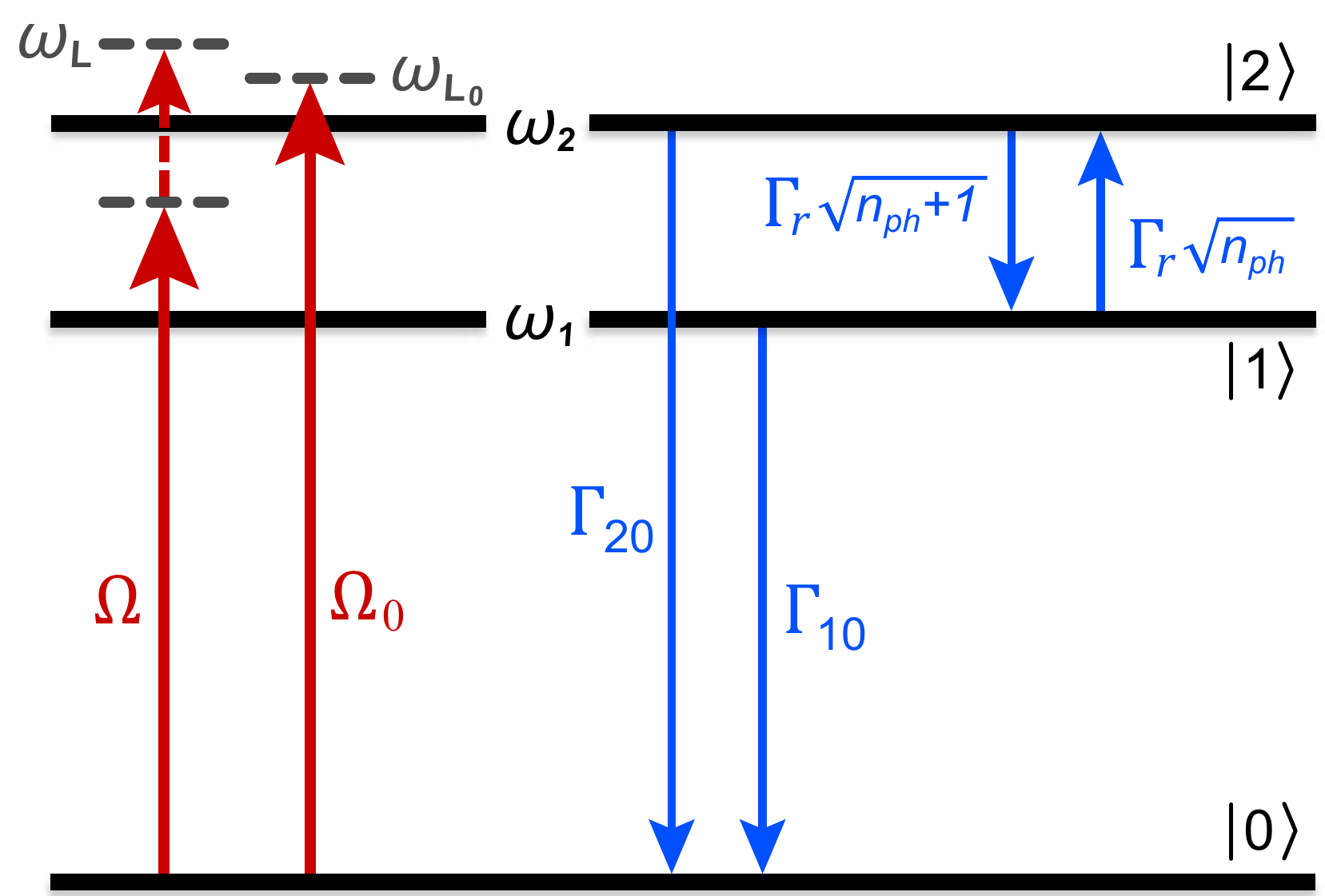}\\
\caption{Energy levels diagram of the system for a down-conversion process in continuous lines. Coherent pumping processes in red arrows labeled $\Omega$ and $\Omega_0$. Dissipative processes in blue arrows  labeled $\Gamma_{10}$, $\Gamma_{20}$, $\Gamma_{r}\sqrt{n_{ph}}$ and $\Gamma_{r}\sqrt{n_{ph}+1}$. Under ideal resonant conditions  $\omega_{L_0}$ coincides with $\omega_2$.}
\label{fig1}
\end{figure}

In this section we develop a theoretical model describing the optical emission properties of our two-QDs simultaneously coupled to a photonic crystal microcavity, under bichromatic resonant excitation. To this end, we consider a 3-level system (Fig.~\ref{fig1}) formed by
\textit{(i)} a ground state labeled $|0\rangle$,
\textit{(ii)} a state labeled $|2\rangle$ to which lasers excite one of the QDs and
\textit{(iii)} a state labeled $|1\rangle$, corresponding to the excitation of either the CM or the other QD, from which the system decays to $|0\rangle$ emitting photons.

Transitions between states $|2\rangle$ and $|1\rangle$ are non-radiative. We study the dynamics of this three-level system upon coherent excitation by two lasers in order to calculate the total intensity of the transition from $|1\rangle$ to $|0\rangle$ (i.e. the population of level $|1\rangle$) as a function of the pumping frequencies and intensities. No distinction is made to whether state $|1\rangle$ corresponds to a QD or the CM. In the latter case, we consider that its population is low enough to be treated as a singly occupied level, as in the case of a QD state. This is a valid approximation since, as revealed by our final results, the decay from  $|1\rangle$ to $|0\rangle$ is much faster than the non-radiative transition from $|2\rangle$ to $|1\rangle$.

The coherent part of the dynamics is described by a hamiltonian (hereafter $\hbar =1$):
\begin{eqnarray}
& H = \omega_1 \sigma_{11} + \omega_2 \sigma_{22} & +\Omega\left(\sigma_{02}e^{i\omega_L t}+\sigma_{20}e^{-i\omega_L t}\right) \nonumber \\
& & +\Omega_0\left(\sigma_{02}e^{i\omega_{L_0} t}+\sigma_{20}e^{-i\omega_{L_0} t} \right)
\end{eqnarray}
where $\omega_1$ and $\omega_2$ are the energies of the excited levels with respect to the ground state, $\Omega$ and $\Omega_0$ are the
Rabi frequencies related to the intensity of the variable and the fixed pump laser, respectively and $\omega_L$ and $\omega_{L_0}$ are their respective energies (cf. Fig.~\ref{fig1}). Operators are defined as $\sigma_{ij} \equiv |i\rangle\langle j|$. In order to simplify the time dependence,
we apply an unitary transformation $U = e^{i\omega_{L_0} t \sigma_{22}}$, giving:
\begin{eqnarray}
& H = \omega_1 \sigma_{11} + \delta_2 \sigma_{22} & +\Omega\left(\sigma_{02}e^{-i\delta_0 t}+\sigma_{20}e^{i\delta_0 t}\right) \nonumber \\
& & +\Omega_0\left(\sigma_{02}+\sigma_{20}\right)
\end{eqnarray}
where $\delta_2 \equiv \omega_2 - \omega_{L_0}$ and $\delta_0 \equiv \omega_{L_0}-\omega_{L} $. The existence of two different excitation energies implies a hamiltonian whose time-dependence cannot be completely
avoided. Apart from this coherent part, there are several dissipative contributions to the dynamics depicted in  Fig.~\ref{fig1}, which are described by Lindblad
terms $\mathcal{L}(\sigma_{ij})=\sigma_{ij}\rho\sigma_{ji}-(\sigma_{jj}\rho+\rho\sigma_{jj})/2$  in a master equation. This includes:

$\bullet $ Radiative decay from $|1\rangle$ to $|0\rangle$: $\Gamma_{10}\,\mathcal{L}(\sigma_{01})$.

$\bullet $ Radiative decay from $|2\rangle$ to $|0\rangle$: $\Gamma_{20}\,\mathcal{L}(\sigma_{02})$.

$\bullet $ Non radiative transition from $|2\rangle$ to $|1\rangle$: $\Gamma_{r}\sqrt{n_{ph}+1}\,\mathcal{L}(\sigma_{12})$.

$\bullet$ Non radiative transition from $|1\rangle$ to $|2\rangle$: $\Gamma_{r}\sqrt{n_{ph}}\,\mathcal{L}(\sigma_{21})$.
	
$\bullet$ Pure dephasing that we consider only for the level under coherent excitation (level $|2\rangle$): $\Gamma_d \, \mathcal{L}(\sigma_{22})$.

$\Gamma_{r}$, $\Gamma_{20}$ and $\Gamma_{10}$ are transition rates, $n_{ph}$ is the phonon population and
$\Gamma_d$ is the pure dephasing rate.
From the master equation we get \cite{Laussy09,delValle09,Sanchez12} a set of Bloch equations for a vector $\vec{\rho} \equiv
(\rho_{22},\rho_{11},\rho_{02},\rho_{20})$ built up with relevant elements of the density matrix $\rho$:
\begin{equation}
\frac{d}{dt}\vec{\rho} = \widehat{M}(t)\vec{\rho} + \vec{P}(t) ,
\label{eq:bloch}
\end{equation}

The components of Eq.~(3)  have the form:
\widetext
\begin{equation}
\widehat{M} = \begin{pmatrix}
-(\Gamma_{20}+\Gamma_{21}) & \Gamma_{12} & -i(e^{it\delta_0}\Omega+\Omega_0)&i(e^{-it\delta_0}\Omega+\Omega_0) \nonumber \\
\Gamma_{21} & -(\Gamma_{10}+\Gamma_{12}) & 0 & 0\\
-2i(e^{-it\delta_0}\Omega+\Omega_0) & -i(e^{-it\delta_0}\Omega+\Omega_0) & -\frac{1}{2}(\Gamma_{20}+\Gamma_{d}+\Gamma_{21}-2i\delta_2) & 0 \nonumber \\
2i(e^{it\delta_0}\Omega+\Omega_0) & i(e^{it\delta_0}\Omega+\Omega_0) & -0 & \frac{1}{2}(\Gamma_{20}+\Gamma_{d}+\Gamma_{21}-2i\delta_2) \nonumber
\end{pmatrix},
\label{eq:equation_too_long}
\end{equation}
\endwidetext
where 

$\Gamma_{12}=\Gamma_r \sqrt{n_{ph}}$, $\Gamma_{21}=\Gamma_r \sqrt{n_{ph}+1}$ 

and

\begin{equation}
\vec{P} = \left[0,0,i(e^{-i t \delta_0}\Omega+\Omega_0),-i(e^{it\delta_0}\Omega+\Omega_0)\right] \nonumber.
\end{equation}

As a time dependence $e^{\pm it\delta_0}$ affects some elements of matrix $\widehat{M}(t)$ and vector $\vec{P}(t)$ in Eq.~(3), one cannot get a steady state $\vec{\rho}_{ss} = -(\widehat{M}^{-1})\vec{P}$. Instead, we make a
Floquet expansion:
\begin{equation}
\vec{\rho} \equiv \sum_{n=0}^{\infty} \vec{\rho}_{n} e^{i n \delta_0 t}
\end{equation}
where $\vec{\rho}_n = \left[\rho_{22}^{(n)}, \rho_{11}^{(n)}, \rho_{02}^{(n)}, \rho_{20}^{(n)}\right]$ \cite{FicekPRA93}. This allows to
separate the matrix $M$ in three terms, $\widehat{M}\equiv \widehat{M}_0 + \widehat{M}_+ + \widehat{M}_-$, where $\widehat{M}_+$ includes
all the terms oscillating with $e^{it\delta_0}$, and $\widehat{M}_-$ all the terms oscillating with $e^{-it\delta_0}$. Similarly, $\vec{P}$
is separated in $\vec{P}_0$, $\vec{P}_+$ and $\vec{P}_-$.
In this framework, the Bloch equations become:
\begin{eqnarray}
i n \vec{\rho}_n - \widehat{M}_0\vec{\rho}_n - \widehat{M}_+\vec{\rho}_{n-1} - \widehat{M}_-\vec{\rho}_{n+1} = \nonumber \\
\vec{P}_0 \delta_{n,0} + \vec{P}_+\delta_{n,1} + \vec{P}_-\delta_{n,-1},
\label{eq:pra}
\end{eqnarray}
which is a system of four equations that can be reduced to a single iterative equation relating the variable of interest $\rho_{11}^{(n)}$ (i.e. the population of the emitter) for different $n$'s.
The iterative equation takes the form of:
\begin{equation}
a_n \rho_{11}^{(n)} + b_n \rho_{11}^{(n-1)} + c_n \rho_{11}^{(n+1)} = d_n ,
\label{eq:recurrence}
\end{equation}
where the coefficients $a_n$, $b_n$, $c_n$ and $d_n$ are given by:
\widetext
\begin{eqnarray}
a_n &\equiv& \frac{\Gamma_{10}+in\delta_0 + \Gamma_{12}}{\Gamma_{21}}(i n \delta_0 + \Gamma_{20}+\Gamma_{21})-\Gamma_{12}-i\Omega_0(F_n^-+F_n^+)-i\Omega G_{n-1}^- - i\Omega G_{n+1}^+ \nonumber \\
b_n &\equiv& -i(\Omega_0 G_n^+ + \Omega F_{n-1}^-) \nonumber \\
c_n &\equiv & -i(\Omega_0 G_n^- + \Omega F_{n+1}^+) \nonumber \\
d_n &\equiv& -i\delta_{n,0}( \Omega_0 C_n^- +\Omega_0 C_n^+ +\Omega D_{n-1}^- + \Omega D_{n+1}^+) -i \delta_{n,1}(\Omega_0 D_n^+ \Omega C_{n-1}^-) \nonumber \\
&& - i\delta_{n,-1}(\Omega_0 D_n^- + \Omega C_{n+1}^+) \nonumber,
\end{eqnarray}
where
\begin{eqnarray}
F_n^\pm &\equiv& \left[in\delta +\frac{1}{2}(\Gamma_{20}+\Gamma_{21}+\Gamma_d\pm 2i \delta_2) \right]^{-1}\, i\Omega_0\ \left( 1+2\frac{\Gamma_{10}+in\delta_0 + \Gamma_{12}}{\Gamma_{21}} \right) \nonumber \\
G_n^\pm &\equiv=& \left[in\delta +\frac{1}{2}(\Gamma_{20}+\Gamma_{21}+\Gamma_d\pm 2i \delta_2) \right]^{-1}\, i\Omega\, \left( 1+2\frac{\Gamma_{10}+i(n \mp 1) \delta_0 + \Gamma_{12}}{\Gamma_{21}} \right) \nonumber \\
C_n^\pm &\equiv& \left[in\delta +\frac{1}{2}(\Gamma_{20}+\Gamma_{21}+\Gamma_d\pm 2i \delta_2) \right]^{-1}\, i\Omega_0 \nonumber\\
D_n^\pm &\equiv& \left[in\delta +\frac{1}{2}(\Gamma_{20}+\Gamma_{21}+\Gamma_d\pm 2i \delta_2) \right]^{-1}\, i\Omega \nonumber.
\end{eqnarray}

\endwidetext
In order to solve this recurrence, we define
vectors $\vec{\rho_{11}}\equiv \left[\rho_{11}^{(-n_S)}, ..., \rho_{11}^{(0)}, ..., \rho_{11}^{(n_S)}
\right]$ and $\vec{d} \equiv \left[d_{-n_S}, ..., d_{0}, ..., d_{n_S} \right]$, where $n_S$ is the number of Floquet satellites we include in the calculation. Then, the solution of Eq.~\eqref{eq:recurrence} takes the form:
$\vec{\rho_{11}} = \widehat{K}^{-1}\vec{d}$, where the elements of the matrix $\widehat{K}$ are
\begin{equation}
K_{nm} = a_n \delta_{nm} + b_n\delta_{m,n-1} + c_n \delta_{m,n+1} .
\end{equation}

Following this approach, the number of satellites included in the expansion is given by the dimension of the matrix $\widehat{K}$. In all cases we have studied, we get convergence of the Floquet expansion for $n_S$=3.

\maketitle

\section{Experimental details}

\begin{figure}[ht]
\includegraphics[width=0.4\textwidth]{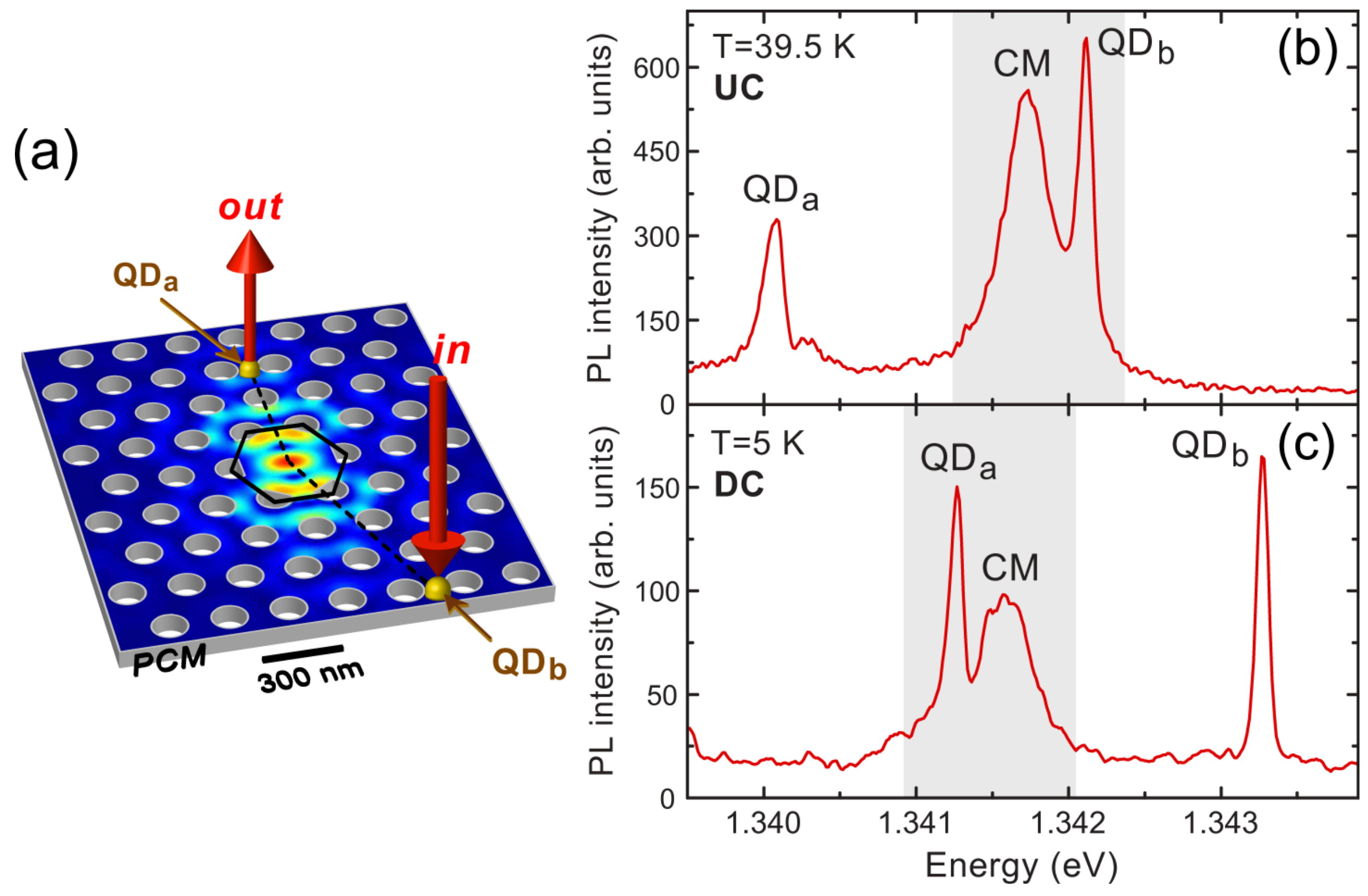}
\caption{\textbf
(a) Sketch of the multi-QD/PCM
system showing the spatial position of the two QDs under investigation (QD$_{\textrm{a}}$ and QD$_{\textrm{b}}$) with respect to the cavity center.
The color intensity plot superimposed on the device structure represents the
electric field pattern of the fundamental cavity mode (CM)
(calculated
using a finite-difference time-domain method). Two-laser excitation (variable and fixed laser) resonant with the QD$_{\textrm{b}}$ transition results in dressed states, which are detected through the emission of QD$_{\textrm{a}}$. Excitation and detection are interchangeable. Photoluminescence (PL) spectra for two different temperatures showing the emission energies of QD$_{\textrm{a}}$, QD$_{\textrm{b}}$ and CM for (b) up-conversion (UC) and (c) down-conversion (DC).}
\label{fig2}
\end{figure}

The measurements were performed on two self assembled InAs QDs, embedded in a photonic crystal microcavity (PCM) (cf. Fig.~\ref{fig2}(a)). The QDs were grown by molecular beam epitaxy inside a $158~nm$ thick GaAs slab on top of a $500~nm$ thick AlGaAs sacrificial layer.
The QD average height and lateral size is $2~nm$ and $50~nm$, respectively. 
The photonic crystal consists of a triangular lattice of air holes
of $140~nm$ diameter with a $230~nm$ pitch, patterned by e-beam lithography and dry etching. An air suspended
membrane was realized by etching of the sacrificial layer. The H1 "calzone" cavity, with a quality factor of about 4000, is
formed by removing the central hole and
modifying the nearest-neighbour holes around the cavity center.
Previous micro-photoluminescence ($\mu~PL$) measurements show that the QDs are located at $0.5\pm 0.15~nm$ and
$0.9\pm 0.15~nm$ away from the cavity mode (CM) center almost in opposite directions \cite{15}, so that the distance between QDs is about $1.4 ~\mu m$, as shown in  Fig.~\ref{fig2}(a). 
Measurements of the exciton spontaneous decay rates of QD$_{\textrm{a}}$ and QD$_{\textrm{b}}$  
as a function of detuning with the CM confirm the existence of Purcell effect and reveal coupling strengths to the cavity of $75~\mu eV$ and $80~\mu eV$, respectively\cite{15}.

Our $\mu~PL$ measurements were carried out using two spatially overlapped Ti-sapphire continuous-wave lasers as fixed and variable excitation sources.
The fixed laser was set at the emission line of one of the QDs and the variable laser was scanned across the same emission line, while the emission intensities of the other QD and the CM were recorded. The two laser beams were spatially overlapped onto a $1.5 \mu~m$ Gaussian spot using a $50 \times $ microscope objective of numerical aperture of $NA = 0.5$, aligned to the center of the PCM. Optical emission, collected by the same objective, was dispersed by a double-grating monochromator of $0.85~m$ focal length and detected by a liquid-nitrogen-cooled charge-coupled camera (CCD). The QDs approximate locations were determined by maximizing their emision intensities under non resonant excitation upon in-plane displacement of the microscope objective by $14 nm$ steps . Measurements were performed in cross-polarized excitation and detection configuration. Partial closing of the intermediate slits of the monochromator allowed the detection of light emitted in a $1~meV$ range as close as $0.5~meV$ from the laser excitation. In this way, the emission intensity of CM and QD$_{\textrm{a}}$ (CM and QD$_{\textrm{b}}$) are recorded  simultaneously in the down-conversion (up conversion) measurements, corresponding to the gray areas in  Fig.~\ref{fig2}. The energy difference between the lowest CM and the emission lines of the two QDs was controlled either by temperature or by deposition of Xe films on the PCM \cite{15}. The detuning between the two QDs was approximately $2~meV$, keeping the CM energy between the QD$_{\textrm{a}}$ and QD$_{\textrm{b}}$ emission lines.

\maketitle
\section{Results and discussion}
Typical PL emission spectra of the coupled QDs-cavity system under non-resonant excitation ($1.41~eV$) are shown in Fig.~\ref{fig2}(b-c) for two temperatures, which determine the energy differences between the CM and the emission form the two QDs.  
Since the emission spectra detected under resonant excitation for both UC and DC experimental configurations (not-shown) are similar to those marked by the gray areas in  Fig.~\ref{fig2}(b) and~\ref{fig2}(c), respectively, the integrated emission intensities of each peak were easily extracted by fitting the recorded spectra to two Gaussians.

Resonant excitation of any of the two QDs produces dressed states due to coherent coupling of the QD to the laser field. We show now that the population of the dressed states of one of the QDs can be read out
through the optical emission of the other QD.

\begin{figure}
\includegraphics[width=0.4\textwidth]{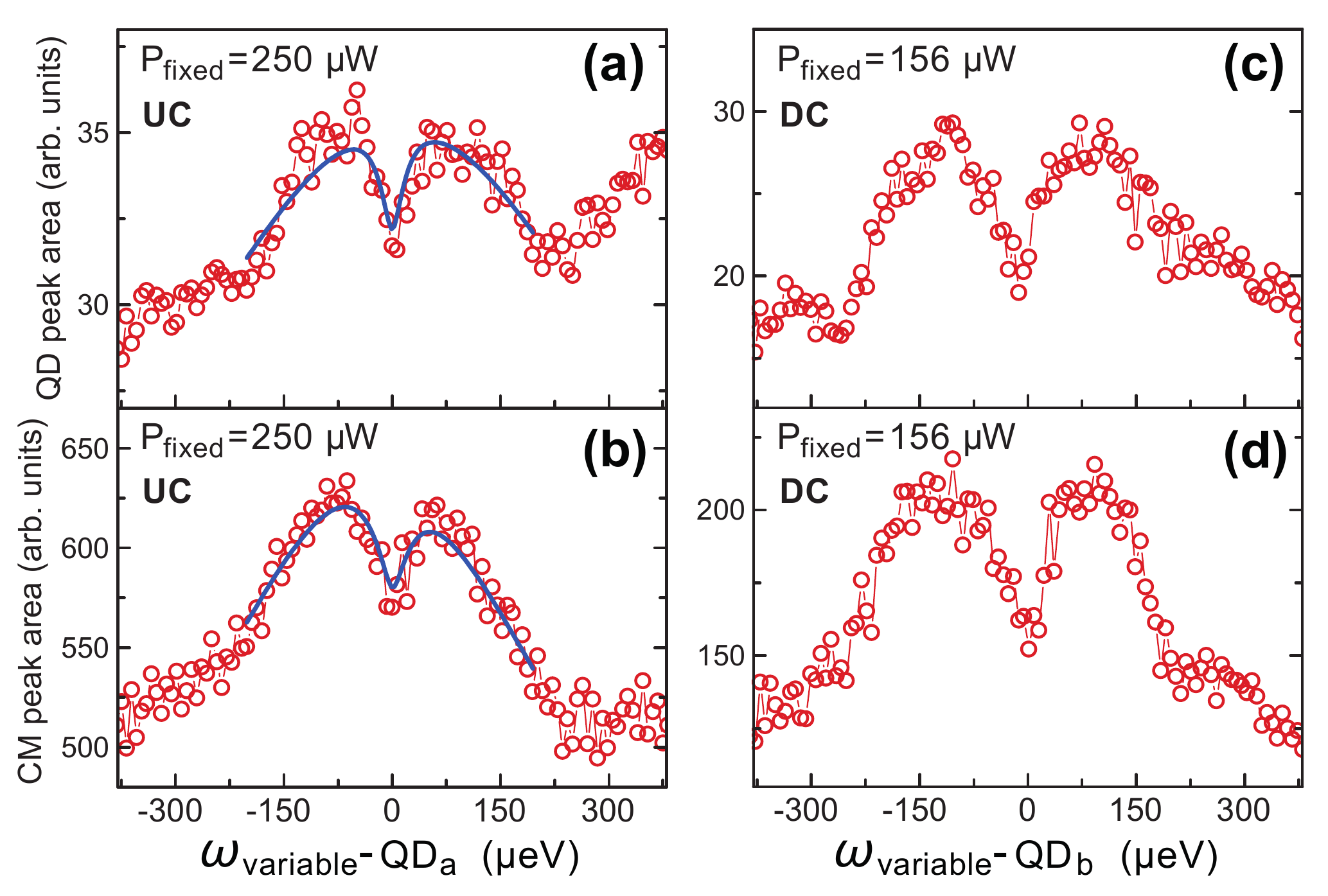}
\caption{Integrated PL peak area of (a) QD$_{\textrm{b}}$ and (b) CM as a function of detuning of the variable laser with respect to the transition energy of
QD$_{\textrm{a}}$.  Same for QD$_{\textrm{b}}$ excitation and (c) QD$_{\textrm{a}}$ and (d) CM detection.}
\label{fig3}
\end{figure}

 The emission intensity of QD$_{\textrm{b}}$ recorded for simultaneous two-laser excitation, as explained in the previous section, is plotted in Fig.~\ref{fig3}(a) as a function of the variable laser detuning with respect to the QD$_{\textrm{a}}$ transition. The fixed laser resonantly exciting QD$_{\textrm{a}}$ has a power intensity of $250~\mu W$, while the variable laser is scanned across the  QD$_{\textrm{a}}$ emission energy with a power of $530~\mu W.$  The measurement is performed in the energy configuration shown in Fig. \ref{fig2}(b). This UC process requires the absorption of acoustic phonons with energy around $2~meV.$
The integrated emission of the CM is shown in Fig.~\ref{fig3}(b) under the same excitation conditions of Fig.~\ref{fig3}(a). Similarly, the integrated emission intensities of QD$_{\textrm{a}}$ and CM are shown in Fig.~\ref{fig3}(c) and~\ref{fig3}(d), respectively, as a function of the variable laser detuning with respect to the QD$_{\textrm{b}}$ transition (DC process).
The "double peak" shape of the QD$_{\textrm{b}}$ integrated emission intensity in Fig.~\ref{fig3}(a) is a consequence of the dressed structure of QD$_{\textrm{a}}$. The first
step in our experiment is
the excitation by two lasers (variable and fixed) of QD$_{\textrm{a}}$, which dresses its quantum states. The second step
is a non-radiative transition to state $|1 \rangle $,
which produces a population of this state
measured in the third step (photon emission).
This double peak is reminiscent of a similar double feature already observed in absorption experiments involving states dressed under the action of two lasers \cite{Wu77, Xu}. The physical origin of this double structure is similar to the Mollow triplet observed in fluorescence experiments except that, in this case, the central feature of the triplet disappears due to a perfect cancellation of absorption and stimulated emission processes \cite{Cohen}. The same spectral shape is obtained when recording the integrated intensity of the CM (Fig.~\ref{fig3}(b)) under the same excitation conditions, as in Ref.~\cite{12} We get an excellent fitting with our model as shown by the continuous lines in Fig.~\ref{fig3}(a) and~\ref{fig3}(b). Fitting parameters are given in the first two rows of Table I. The main difference of fitting parameters between Figs.~\ref{fig3}(a) and~\ref{fig3}(b) is the overall intensity F, as well as $\Gamma _r$, which is weaker in the case of QD$_{\textrm{b}}$ readout. The rest of the parameters have values with moderate changes within the fitting uncertainty. In particular, the small asymmetry change between Figs.~\ref{fig3}(a) and~\ref{fig3}(b) is due to a slight accidental detuning of the fixed laser with respect to the QD$_{\textrm{a}}$ emission \cite{12} ($\delta _2$ in our model). This detuning is of the order of $\ 30~\mu eV$, i.e. one order of magnitude smaller than the width of the QD PL-emission (cf. Fig. \ref{fig2}). 

The behavior of the total emitted intensity in the DC process (Figs.~\ref{fig3}(c) and~\ref{fig3}(d)) is qualitatively similar to the UC case, although fits by the model are not numerically stable, thus reducing their reliability. The reason lies probably in the weaker and noisier signal in the DC spectra as compared to those of the UC ones.This prevents an overall fitting of all experimental data with common parameters. Nevertheless, Fig.~\ref{fig3}(c) indicates that the spectral distribution of the dressed states of the excited QD$_{\textrm{b}}$ is translated to the energy dependence of the integrated emission intensity of the "detector" QD$_{\textrm{a}}$.

\begin{figure}[ht]
\centering
\includegraphics[width=0.4\textwidth]{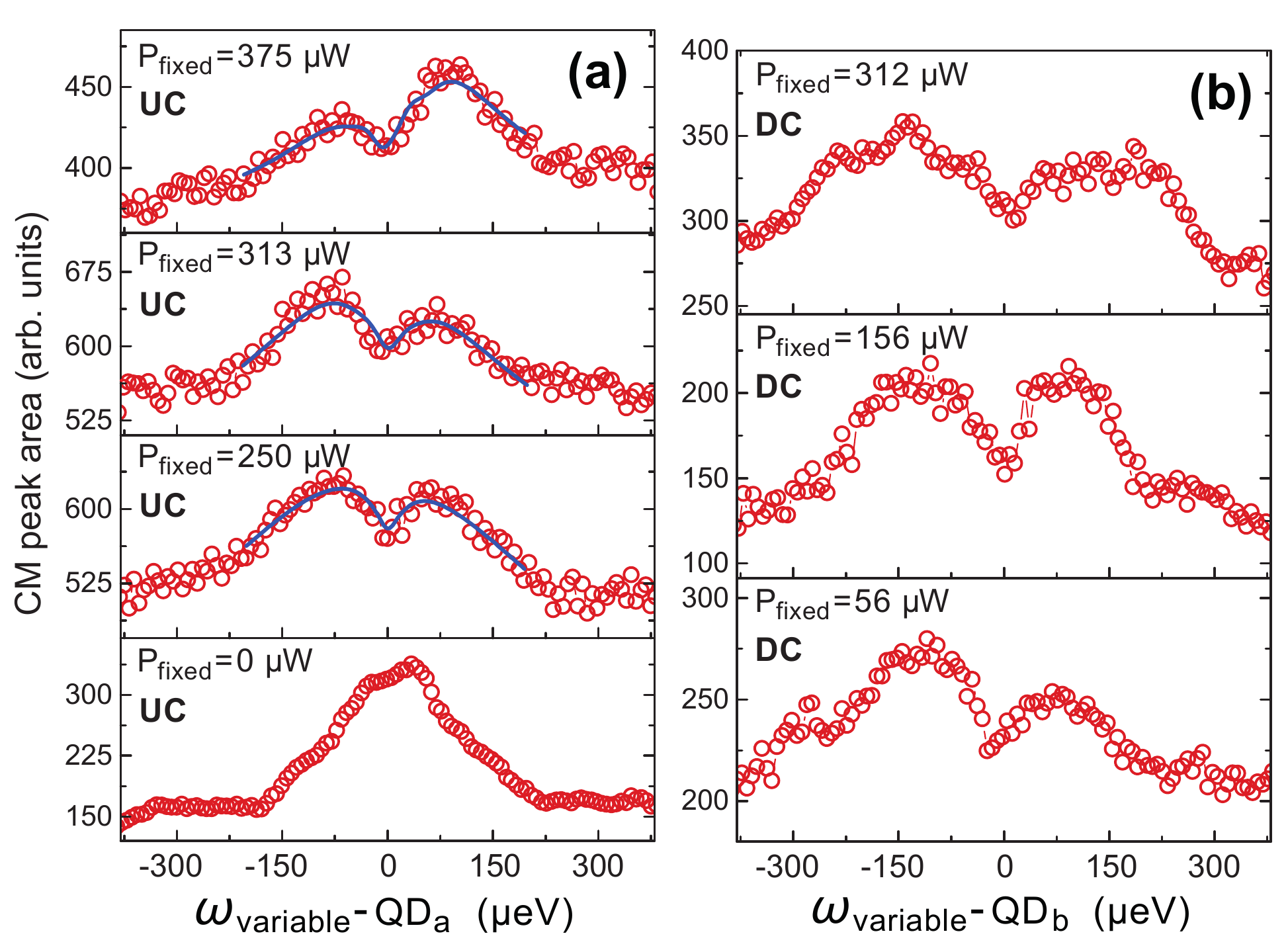}
\caption{Integrated PL peak area of the CM as a function of
detuning of the variable laser with respect to the transition energy of (a) QD$_{\textrm{a}}$ and (b) QD$_{\textrm{b}}$.}
\label{fig4}
\end{figure}

Our model also describes the dependence of the dressed state spectral distribution on excitation power of both fixed and variable
laser. It must be underlined that we do not expect a
square root dependence of the splitting on the fixed laser power, because both fixed and variable laser intensities are comparable. Instead, the energy distribution of the dressed state population
depends on the combined action of the two lasers in a non-trivial way.

Figure~\ref{fig4} shows power dependent measurements
of the cavity emission, as it is stronger than that of the QDs. The integrated emission intensity of the CM for different excitation intensities ($P_{fix} \propto \Omega_0$) of the fixed laser is shown in Fig.~\ref{fig4}(a) for the UC case and in  Fig.~\ref{fig4}(b) for the DC case. For excitation with the variable laser alone (bottom panel), a single peak is obtained
evidencing the cavity-QD coupling upon excitation resonant with the s-states. Its width ($205~\mu eV$), however, is $30 \% $ higher than the width of the non-resonantly excited QD$_{\textrm{a}}$ emission ($160~\mu eV$) shown in Fig.~\ref{fig2}. This broadening reflects the relaxation process between states $|2\rangle$ and $|1\rangle$ (cf. Fig.~\ref{fig1}), which is absent in a pure absorption or emission transition. The solid lines again correspond to fits produced by our theoretical model, whose fitting parameters are shown in Table I (rows 2 to 4).  The parameters listed in Table I correspond to a joint best fit of spectra in Figs.~\ref{fig3} and~\ref{fig4}. The decrease of the dephasing rate with pumping power observed in Table I suggests that inhomogeneous broadening in our experiment is not only due to noise produced by phonons, but also to other mechanisms as input/output coupling efficiency \cite{Loo}. Note, that the coupling strengths $\Gamma_{10}$ and $\Gamma_{20}$ do not change upon increasing $\Omega_0$ while $\Gamma_r \sqrt{n_{ph}}$ increases significantly with pump power. This is partially attributed to an increase in the phonon occupation number $n_{ph}$.

\section{Summary}
In summary, we demonstrate dressing of excitonic QD states by two-colour resonant excitation in a system formed by two QDs weakly coupled to a common cavity mode. The spectral distribution of the dressed states of any of the QDs is efficiently readout by the integrated emission intensity of the other QD, which is spatially separated by $1.4~\mu m$ from the excited one, as well as by the cavity emission intensity. The experimental results are explained by a theoretical model describing the dynamics of a three level system coherently excited by the two lasers. The efficient excitation and detection through different spectral channels, which are interchangeable, demonstrates the feasibility for the use of QDs as nodes in an integrated quantum network based on coupled multiple QDs-cavity systems, even when the two QDs are not identical.

\widetext

\begin{table}
\begin{tabular}{|c|c|c|c|c|c|c|c|c|c|}

  \hline
  & $\Omega_0 (\mu eV) $& $\Omega/\Omega_0 $&$ \Gamma _{10}(\mu eV)$ & $\Gamma_{20}(\mu eV)$ &
  $\Gamma_r \sqrt{n} (\mu eV)$ & $\Gamma_d (\mu eV)$ & $\delta _2 (\mu eV)$ & F & Fig. \\
 \hline
 QD$_{\textrm{b}}$ & 15.4 & 2.12 & 108 & 7.0 & 0.08 & 616 & -9 & 1369 & 3(a) \\
 CM & 15.4 & 2.12 & 103 & 7.1 & 0.15 & 437 & 20.8 & 7000 & 3(b),4(a)\\
 CM & 20.0 & 1.7 & 105 & 7.1 & 0.30 & 281 & 18.2 & 4400 & 4(a)  \\  
 CM & 23.5 & 1.4 & 101 & 7.1 & 0.38 & 187 & -27.8 & 2500 & 4(a)  \\
  \hline
\end{tabular}
\caption{Parameters used for the fittings in Figs.~\ref{fig3}(a,b) and~\ref{fig4}(a). Note that data in Fig.~\ref{fig3}(b) and Fig.~\ref{fig4}(a)-third panel are the same.}
\end{table}
\endwidetext


\textbf
This work was supported by the Spanish MINECO under contract MAT2011-22997, by CAM under contract S2009/ESP-1503 and by the FP7 ITN Spin-optronics (237252). C.S-M. and E.C. acknowledge FPI grants.

\end{document}